\documentclass[journal]{new-aiaa}
\usepackage{amsmath}
\usepackage[version=4]{mhchem}
\usepackage[nameinlink, capitalize]{cleveref}
\usepackage[version=4]{mhchem}
\usepackage[most]{tcolorbox}
\usepackage{nomencl}
\usepackage{doi}

\definecolor{myCol}{rgb}{0.9, 0.9, 0.9}

\newcommand{\wt}[1]{{\widetilde{#1}}}
\newenvironment{itemizePacked}{
\begin{itemize}
  \setlength{\itemsep}{1pt}
  \setlength{\parskip}{0pt}
  \setlength{\parsep}{0pt}
}{\end{itemize}}
\def\cE{{\cal{E}}}
\def\cC{{\cal{C}}}
\def\p{\partial}
\def\phivec{{\mbox{\boldmath$\phi$}}}
\def\uvec{{\mbox{\boldmath$u$}}}
\def\omegavec{{\mbox{\boldmath$\omega$}}}
\def\xivec{{\mbox{\boldmath$\xi$}}}
\def\xvec{{\mbox{\boldmath$x$}}}
\def\varphivec{{\mbox{\boldmath$\varphi$}}}
\def\cM{{\cal{M}}}
\def\psivec{{\mbox{\boldmath$\psi$}}}
\def\d{\textup d}

\hypersetup{colorlinks=true}
\title{Requirements Towards Predictive Simulations of Turbulent Reacting Flows}
\author{Matthias Ihme\footnote{mihme@stanford.edu}}
\affil{Department of Mechanical Engineering, Stanford University, Stanford, CA 94305}
\begin{document}
\maketitle
\begin{abstract}
Significant progress has been made on the model development for simulating turbulent reacting flows. As a consequence, we are currently in a position where key-physical aspects of fairly complex combustion processes are well understood at a qualitative and -- in many cases --  also at a quantitative level. Examples are the prediction of temperature and major species, statistically stationary flames, gas-phase combustion, turbulent transport, and turbulence/flame coupling. However, current  challenges lie in capturing transient processes and stability boundaries, minor species and emissions, multiphase flows and phase-transition, as well as multidimensional flame/flow interactions that may involve flame curvature effects, stratification, partially premixing, or flame-wall interaction. With this, the question arises what steps need to be taken to elevate the current state of modeling capabilities in order to address these deficiencies? This paper seeks to address this question. We begin by reviewing the current state of combustion model approaches, our quest for improving existing models, the separation of errors arising from numerical discretization and physical models, and ideas on model evaluations. We then proceed by examining concepts on quantitative model evaluations, requirements on predictability, quantities of interest, and cost/accuracy trade-offs. We close by introducing recent concepts  that assimilate time-resolved measurements into numerical simulations for state estimation, model evaluation, and parameter determination.
\end{abstract}
\section{Introduction}
The objective of this paper is to discuss research needs and requirements towards the predictive simulation of turbulent reacting flows. While significant progress has been made on the fundamental understanding of turbulent combustion, the modeling of stationary combustion processes and the qualitative evaluation of combustion dynamics, deficiencies remain in regard to quantitative predictions and reliable simulations of combustion processes, involving  complex geometries, unsteady operating conditions, pollutants, and the consideration of multiphase flows. Faced with these challenges, questions arise on how to select a particular simulation approach to predict turbulent reacting flows under specific constraints about solution accuracy and computational resource allocation, and what are techniques that allow us to assess the accuracy of a turbulent-combustion simulation.  

In the following, we consider simulation methods for modeling gaseous turbulent combustion, consisting of the following components: a discretized representation of the flow geometry; initial and boundary conditions; a mathematical model for the numerical solution of the governing equations, such as Reynolds-averaged Navier-Stokes (RANS), large-eddy simulation (LES), and direct numerical simulation (DNS);  a combustion model for the representation of energy release, species conversion and the coupling to density, pressure, and constitutive properties; closure models for the description of the feedback of turbulence with the gas-phase combustion; a chemical-kinetic mechanism; and constitutive relations for the thermo-chemical quantities. Furthermore, the selection of a particular simulation approach is constrained by user-specific requirements on the accuracy in predicting certain responses functions (RFs), comprising quantities of interest (QoIs) or processes of interest (PoIs). These requirements have to be considered in the presence of computational resources that impose limitations on the choice of certain simulation approaches and modeling strategies. 

In the following, we seek to address several questions with the goal of examining the current state-of-the-art in combustion modeling, identifying research gaps, and stipulating   ideas to overcome these challenges. Specifically,~\cref{SEC_SIM_PURP} is concerned with the selection of a simulation approach that aligns with the purpose of a simulation. We then proceed by discussing the selection of combustion models and discuss different modeling strategies to represent turbulent combustion. Aspects on the importance of numerical errors and the interaction of these errors with physical models are discussed in~\cref{SEC_NUM_DISCR}.~\Cref{SEC_MODEL_ACCURACY} is concerned with the qualitative assessment of the simulation accuracy, and recent developments on the probabilistic analysis and dynamic simulation content are discussed. Emerging concepts for augmenting simulations with experimental data is addressed in~\cref{SEC_AUG_SIM}, and the paper finishes with a summary in~\cref{SEC_CONCL}.
\section{\label{SEC_SIM_PURP}What is the purpose of a combustion simulation?}
\begin{figure}[!htb!]
  \centering
\includegraphics[width=0.6\textwidth, clip=, keepaspectratio]{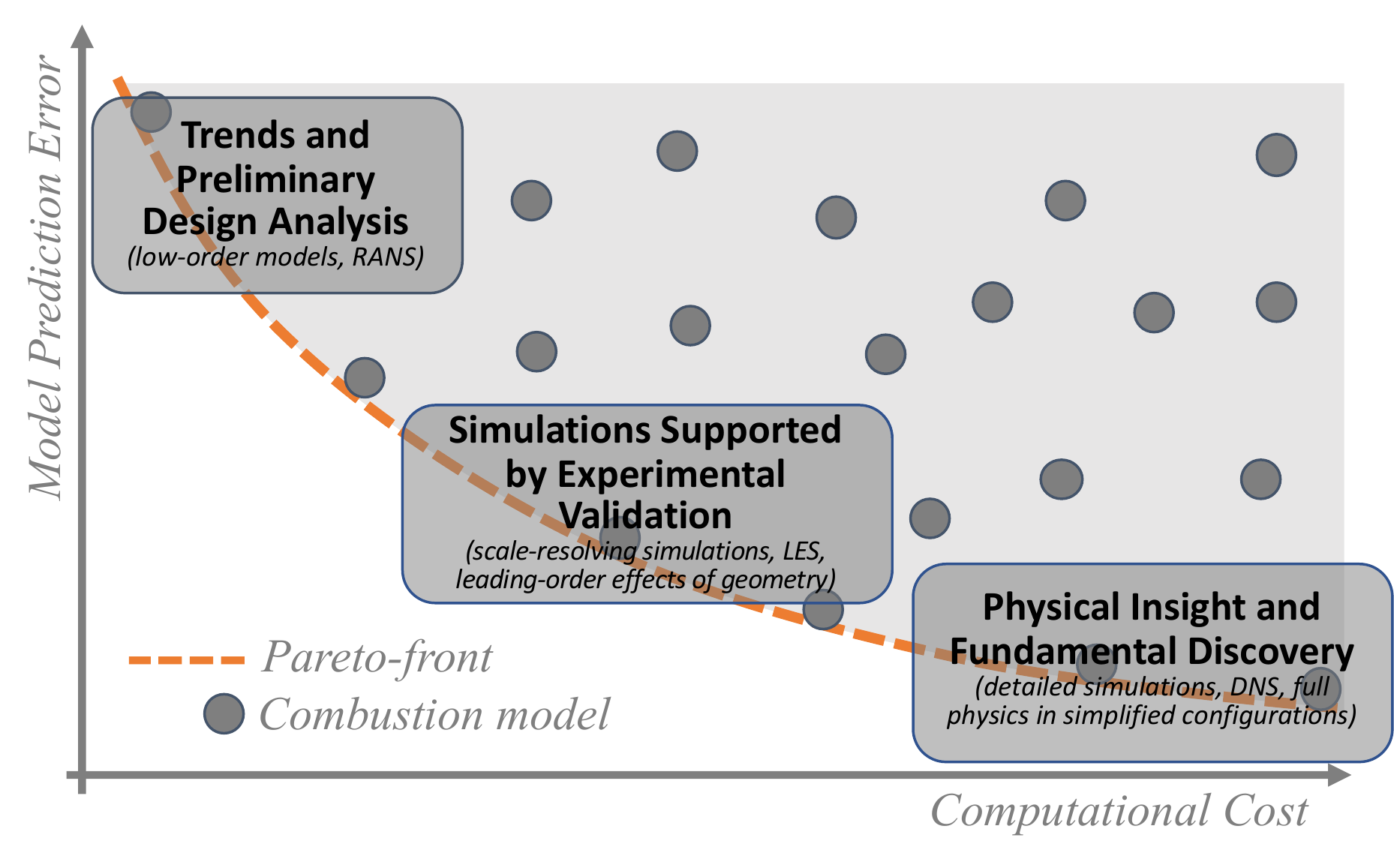}
\caption{\label{FIG_MODEL_PERFORMANCE}Schematic illustrating the selection of simulation approach and dependence on requirements about accuracy and computational cost.}
\end{figure}
The first and perhaps most important question to ask is in regard to the objective of a particular combustion simulation. The answer to this questions  will depend on the field of application (see~\cref{FIG_MODEL_PERFORMANCE}): while industrial applications might utilize simulation tools to support the design optimization with the goal of minimizing emissions or exploring new combustion strategies, academia and research laboratories might employ simulations for gaining principal understanding about flame-structure, energy transfer or other fundamental combustion-physical processes. Other applications may rely on simulations to support the diagnosis of failure modes or for risk assessment. As such, it becomes apparent that the specific purpose determines the choice of a particular simulation approach, the model selection, as well as requirements on the accuracy of the simulation and computational resource allocation.

Most common is the prediction of certain scalar response quantities, such as total heat release, fuel conversion, temperature, or pollutants. These response quantities  are either directly solved for in a simulation or can be determined by post-processing the simulation results. Results are commonly reported in the form of statistical quantities, ensemble-averaged quantities, or conditional data in order to facilitate analysis or comparison with measurements. Temperature, species of \ce{CO2} and \ce{H2O}, and heat release, which is controlled by few exothermic reaction pathways, can all be predicted to good accuracy independent of a particular combustion regime~\cite{WU_SEE_WANG_IHME_CF2015,SEE_IHME_PCI2015}. This, however, is not the case for minor species and pollutants such as CO, \ce{NO_x}, or soot. Since these quantities are tightly regulated in aviation gas turbines, stationary power generation, and automotive applications, their reliable prediction introduces significant modeling challenges. 

In contrast to predicting QoIs, the simulation of transient processes becomes relevant for determining combustion dynamics, such as ignition, blow-out limits, stability boundaries, onset of engine-knock conditions, extinction, or thermoacoustic instabilities. Since these processes of interest (PoIs) are deterministic chaotic phenomena that are characterized by aperiodic long-term behavior with strong sensitivity to initial conditions, probabilistic techniques are most likely more suitable to describe these events.

The purpose of a simulation is linked to specific requirements on solution accuracy, model fidelity, and resource allocation, as illustrated in~\cref{FIG_MODEL_PERFORMANCE}. For instance, the computational exploration of a high-dimensional parameter space to optimize the burner performance emphasizes the need for low-order models at the expense of solution accuracy, and targeted simulations of higher fidelity are then conducted to validate the most promising design candidates. Time-to-solution and computational resource allocation are constraining factors on computational simulations. 
While the computational cost can be estimated reasonably well from the knowledge about the algorithmic complexity, spatio-temporal resolution, and code-scalability, quantifying improvements in solution accuracy of a specific model prior to performing the simulation is not trivial.

\begin{tcolorbox}[colback=myCol,breakable]
Recommendations:
\begin{itemizePacked}
\item Establish formal methods for down-selecting specific simulation approaches in order to meet user-specific requirements of combustion simulations.
\item Articulate specific requirements on solution accuracy, response functions, and resource allocations to obtain improvements in solution accuracy $\cE$ for certain computational cost $\cC$.
\end{itemizePacked}
\end{tcolorbox}
\nomenclature{$\cE$}{error in solution accuracy}
\nomenclature{$\cC$}{computational cost}
\section{\label{SEC_COMB_MODEL}How should a combustion model be selected?}
The computational modeling of turbulent reacting flows is concerned with overcoming challenges of resolving all relevant spatio-temporal scales and the consideration of a large number of chemical species that participate in the combustion process. The evolution of the chemical species can be obtained as solution to a advection-diffusion-reaction equation, which is here written in general form:
\begin{equation}
\label{EQ_SPECIES_TE}
 \p_t (\rho \phivec) + \nabla\cdot(\rho\uvec\phivec) = -\nabla\cdot{\bf j}_{{\phi}}+\rho\dot{\omegavec}_{{\phi}}\;,
\end{equation}
where $\rho$ is the density, $\uvec\in\mathbb{R}^{N_d}$ is the velocity vector, $\phivec\in\mathbb{R}^{N_s}$ is the vector of species mass fractions, ${\bf{j}}_{{\phi}}\in\mathbb{R}^{N_s\times N_d}$ is the mass-diffusion flux matrix, and $\dot{\omegavec}_{{\phi}}\in\mathbb{R}^{N_s}$ is the vector of reaction rates. 
\nomenclature{$\rho$}{density}
\nomenclature{$\uvec$}{velocity vector}
\nomenclature{$\phivec$}{vector of species mass fractions}
\nomenclature{${\bf{j}}_{\zeta}$}{diffusion flux matrix of quantity $\zeta$}
\nomenclature{$\dot{\omegavec}$}{vector of reaction rates}
\nomenclature{$\wt{(\cdot)}$}{Favre-filtered quantity}
\nomenclature{$\overline{(\cdot)}$}{Reynolds-filtered quantity}

To overcome the challenge of describing effects that occur at computationally unresolved scales, involving turbulent stresses, turbulent scalar transport, and turbulence/chemistry interaction,~\cref{EQ_SPECIES_TE} is filtered or averaged using a Favre-filtering procedure:
\begin{equation}
\label{EQ_SPECIES_TE_FILTER}
 \p_t (\overline{\rho} \wt{\phivec}) + \nabla\cdot(\overline{\rho}\wt{\uvec}\wt{\phivec}) = -\nabla\cdot\overline{\bf j}_{{\phi}}-\nabla\cdot{\bf j}_{{\phi}}^{\text{sgs}}+\overline{\rho}\wt{\dot{\omegavec}}_{{\phi}}\;,
\end{equation}
which introduces unclosed terms for the subgrid scalar flux ${\bf j}_{\phi}^{\text{sgs}}$ and the filtered chemical source term $\wt{\dot{\omegavec}}_{{\phi}}$. These terms require modeling. Here, the low-pass filtering follows the common definition of a Favre-filter with $\wt{\phi}=\overline{\rho}^{-1}\int \rho(\xivec,t)\phi(\xivec,t)G(\xvec-\xivec,t) \d\xivec$ with $G$ being the compact filter kernel. Closure models for ${\bf j}_{\phi}^{\text{sgs}}$ are commonly adopted from non-reacting and isothermal flow models and corrections are introduced to account for variable transport properties and density variations. 

To reduce the chemical complexity,~\cref{EQ_SPECIES_TE_FILTER} is replaced by transport equations associated with appropriate low-dimensional manifold methods, in which the chemical state vector is represented in terms of a low-dimensional manifold:
\begin{equation}
 \phivec\simeq{\varphivec} = \cM(\psivec)\;,
\end{equation}
\nomenclature{$t$}{time}
\nomenclature{$\psivec$}{state vector describing combustion manifold}
\nomenclature{$\cM$}{low-dimensional manifold}
\nomenclature{sgs}{subgrid-scale quantity}
\nomenclature{$N$}{number}
\nomenclature{$T$}{temperature}
\nomenclature{$Z$}{mixture fraction}
where $\psivec\in\mathbb{R}^{N_\cM}$ is the state vector that is used to parameterize the manifold. The vector $\psivec$ may include a subset of species mass fractions, mixture fraction, reaction progress, as well as other flow-field describing quantities, such as strain rate or scalar dissipation. The topology of the manifold and with this the manifold-spanning quantities depend on a particular combustion model. The evolution of the manifold-spanning state-vector $\psivec$ is then obtained as solution to a transport equation:
\begin{equation}
\label{EQ_SPECIES_TE_FILTER_MFOLD}
 \p_t (\overline{\rho} \wt{\psivec}) + \nabla\cdot(\overline{\rho}\wt{\uvec}\wt{\psivec}) = -\nabla\cdot\overline{\bf j}_{{\psi}}-\nabla\cdot{\bf j}_{{\psi}}^{\text{sgs}}+\overline{\rho}\wt{\dot{\omegavec}}_{{\psi}}\;.
\end{equation}
Therefore, by using the manifold formulation, the number of equations that is solved reduces from $N_s$ to $N_\cM$. In the case that $N_\cM=N_s,$ the manifold model recovers the chemical complexity of the complete chemical system.

Over recent years, different modeling approaches have been developed that address the reduction of the chemical complexity and the modeling of the turbulence/chemistry interaction. At a fundamental level, these approaches can be distinguished into \emph{topology-free} and \emph{topology-based} combustion models~\cite{WU_SEE_WANG_IHME_CF2015}. Examples of topology-free combustion models are finite-rate chemistry models, the eddy-dissipation concept~\cite{MAGNUSSEN_HJERTBERG_SYMP16}, probability-density function (PDF) methods~\cite{POPE_PECS85,HAWORTH_PECS2010}, and deconvolution methods~\cite{PANTANO_SARKAR_POF2001,DOMINGO_VERVISCH_PCI2014,WANG_IHME_CF2017,WANG_IHME_CF2019}. These combustion models introduce limited assumptions about the flame structure, and are therefore considered to be applicable to a wide range of combustion problems. Topology-free models enable the incorporation of different combustion-physical processes, but are accompanied by higher  computational complexity. In contrast, topology-based combustion models exploit the topological structure of the flame by constructing the manifold from the solution of canonical flame configurations, such as  laminar counterflow diffusion flames, freely propagating premixed flames, or one-dimensional embedded flame elements. Examples of topology-based combustion models are the class of flamelet models~\cite{PETERS_PECS84,GICQUEL_DARABIHA_THEVENIN_PCI28,OIJEN_DEGOEY_CST2000,CHUCK_JFM04,IHME_PCI04}, thickened flame models~\cite{COLIN_VEYNANTE_POINSOT_POF2000}, and conditional moment closures~\cite{KLIMENKO_BILGER_PECS99,KLIMENKO_CTM01}. 

\begin{figure}[!htb!]
\centering
\includegraphics[width=0.5\textwidth, clip=, keepaspectratio]{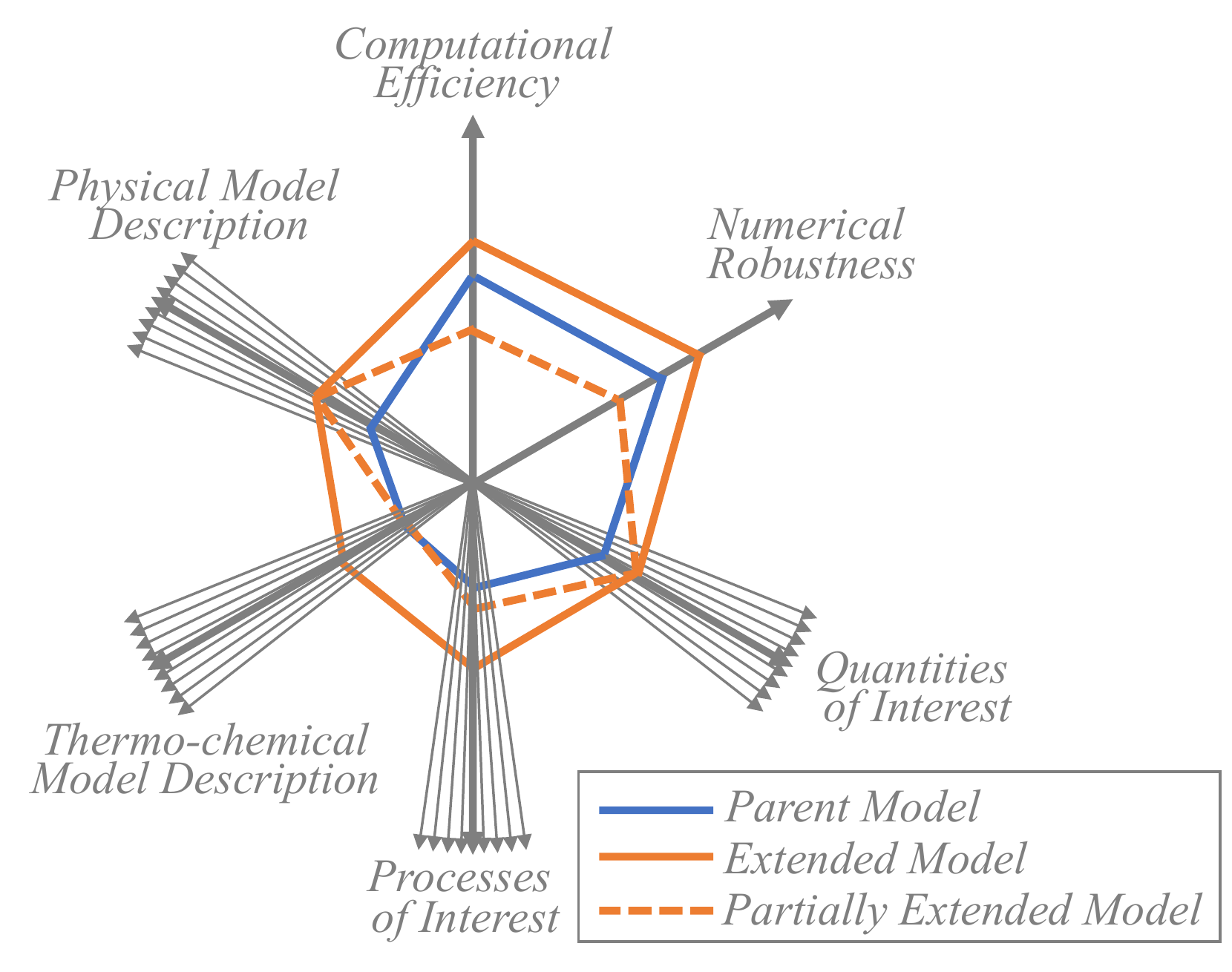}
\caption{\label{DG_MODEL_EXTENSION}Radar chart, schematically illustrating the model performance and impact of partial and complete model extension. Directions along arrow indicates improved model performance, multiple thin arrows emphasize that several processes or quantities of interest require consideration.}
\end{figure}

Common to all models are assumptions that are invoked to make them tractable. These assumptions include the consideration of particular combustion modes, transport properties, chemical complexity, contraction of chemical time-scales, or the omission of multidimensional, non-equilibrium and radiation effects. These models have been examined through \emph{a priori} analyses or \emph{a posteriori} comparisons with experimental observations and high-resolution simulation data. Model extensions have been proposed to address shortcomings that are encountered in applications to new flame configurations, combustion regimes, or operating conditions. However, in these extensions it is critical to ensure that the parent model constitutes a proper subset of the extended model in order to recover the performance of the underlying model (see~\cref{DG_MODEL_EXTENSION}). Examples of model extensions are the consideration of heat-loss effects~\cite{MA_WU_IHME_HICKEY_AIAAJ2018,FIORINA_CTM03}, radiation~\cite{IHME_PITSCH_POF2008}, multistream flows~\cite{HASSE_PETERS_PCI30,IHME_SEE_PCI33,IHME_ZHANG_HE_DALLY_FTC2012,CHEN_IHME_CF2013}, and transient processes~\cite{IHME_SEE_CF2010}. While these model extensions fulfill the requirement on the subset-completeness, increasing storage requirements of these higher-dimensional manifold representations can impact the numerical accuracy and may challenge the resolution requirements in accurately representing the combustion manifold. Furthermore, the solution of additional transport equations that often involve high-order moments introduce additional challenges in obtaining stable solutions and the modeling of unclosed terms. As such, it becomes clear that further extensions of manifold-based combustion models will face limitations. In light of rapidly increasing computational resources, efficient time-integration schemes with linear complexity~\cite{WU_MA_IHME_CPC2019}, efficient chemical reduction techniques~\cite{LU_LAW_PCI30,PEPIOT_PITSCH_CF2008,NIEMEYER_SUNG_RAJU_CF2010,JARAVEL_WU_IHME_CF2019}, and the development of realizable reconstruction techniques for subgrid contributions~\cite{WANG_IHME_CF2017,WANG_IHME_CF2019,NIKOLAOU_CANT_VERVISCH_PRF018}, it is expected that the utilize of topology-based combustion models will decline in the foreseeable future.

\begin{figure}[!htb!]
\centering
\includegraphics[width=0.71\textwidth, clip=, keepaspectratio]{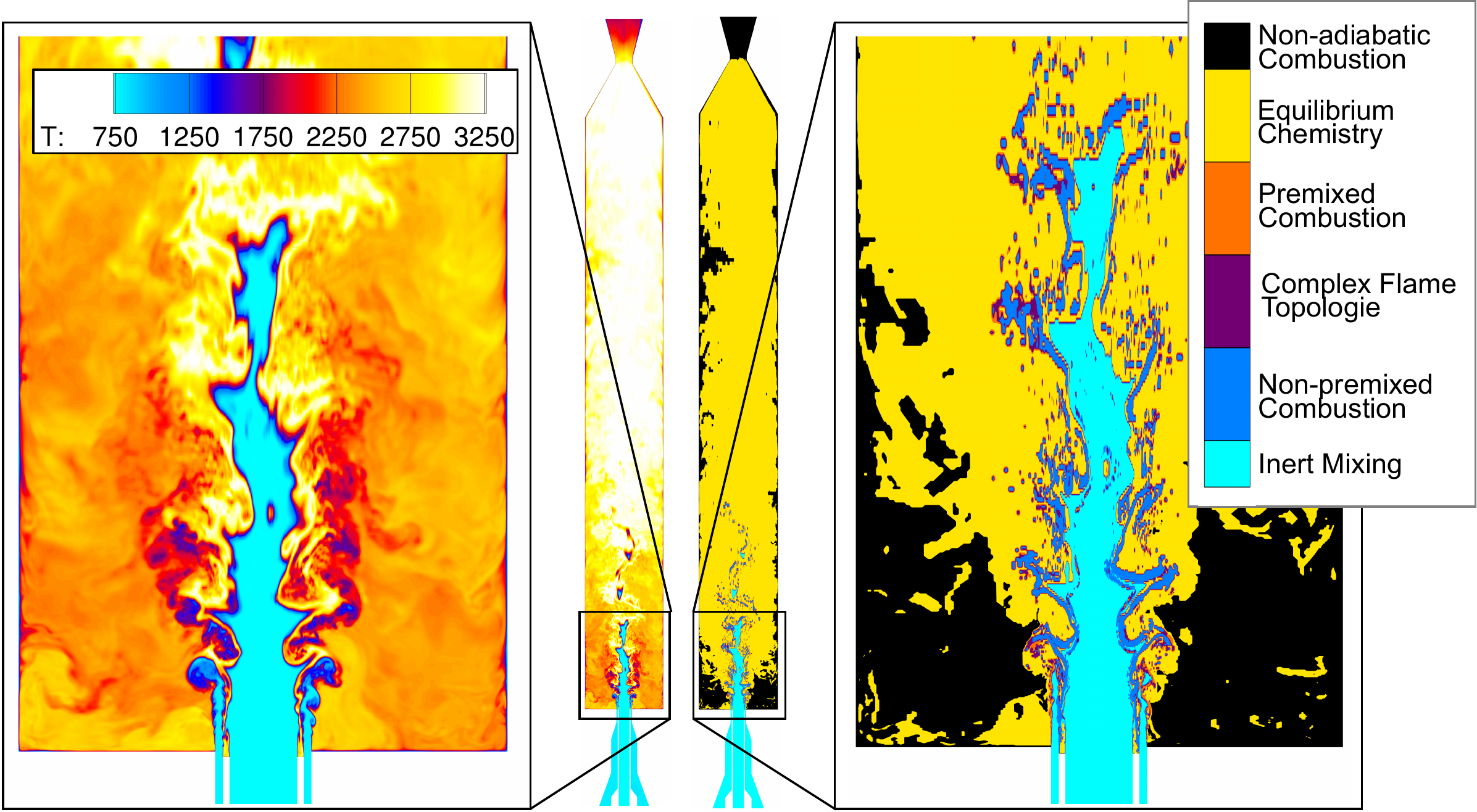}
\caption{\label{FIG_COMB_MODE}Simulation of uni-element rocket injector~\cite{MA_WU_IHME_HICKEY_AIAAJ2018}, showing instantaneous temperature field (left) and combustion modes (right).}
\end{figure}

Combustion models are commonly employed in a \emph{monolithic} form, meaning that only a single model is utilized to describe the entire combustion process encountered in a combustor configuration. While this is certainly appropriate for flames that operate in the vicinity of asymptotic regimes of premixed, non-premixed, or near-equilibrium conditions, these monolithic modeling strategies introduce challenges for representing complex combustion processes involving multimode combustion, multiphase flows, or the consideration of localized combustion events such as flame/wall interaction, ignition, or flame/turbulence coupling. This is illustrated in~\cref{FIG_COMB_MODE}, showing instantaneous simulation results of a single-element coaxial rocket injector~\cite{MA_WU_IHME_HICKEY_AIAAJ2018}. The instantaneous temperature field is shown in the left panel and combustion relevant combustion regimes are presented on the right, emphasizing the complexity of turbulent combustion.

To reduce the modeling complexity, opportunities arise for combining different combustion models in such a way as to optimally represent specific thermochemical processes that are encountered in a combustion simulation. This principle is encapsulated in the Pareto-Efficient Combustion (PEC) framework~\cite{WU_SEE_WANG_IHME_CF2015,WU_MA_JARAVEL_IHME_PCI2019}, in which an optimal combustion-submodel assignment is employed to meet user-specific requirements about solution accuracy on specific QoIs under consideration of computational-cost constraints. This submodel assignment takes advantage of existing combustion models that are readily available in existing CFD-solvers. PEC is a trust-region formulation and utilizes a \emph{Pareto efficiency} to facilitate an optimal combustion submodel assignment. The notion of Pareto efficiency describes an optimality between two competing conditions -- in the PEC model these two conditions are (i) the accuracy in predicting user-specific quantities of interest and (ii) computational cost. Pareto-optimality is then achieved when it is impossible to make any one condition better without making the other condition worse. In the present case, this results in a Pareto front. 

By combining different combustion models, PEC enables the general model-adaptation to the underlying flow-field representation so that regions of different combustion-physical complexities are represented by the most appropriate model without violating intrinsic model assumptions. Key attributes of this PEC-formulation are (i) the local adaptation of the model fidelity and computational complexity to the underlying flow-field-specific combustion processes, (ii) the control of the model accuracy and computational cost through the selection of error threshold, model compliance, and quantities of interest, and (iii) the representation of realistic fuel chemistry through detailed and reduced kinetic models. To enable the application to turbulent combustion requires  the development of algorithms that take into consideration a robust model selection, error-evaluation, subzone coupling, conservation properties, and considerations in regard to load-balancing and dynamic submodel adaptation. 

In this context, we note that other adaptation techniques have been developed, which are concerned with locally adapting the chemical complexity to thermodynamic activity in flow simulation~\cite{LIANG2009527,SHI_LIANG_GE_REITZ_CTM2010,XU_AMEEN_SOM_CHEN_REN_LU_CF2018,
GOU2013225,
HE_ETAL_EF2011,
REN201419,
BANERJEE2006619,
LIANG20153236,
GREEN_BARTON_ETAL_IECR2001,
SCHWER2003451}. As such, PEC generalizes these dynamic adaptive chemistry techniques by adapting the combustion-model assignment to meet user-specific requirements about solution-accuracy with respect to user-selected quantities of interest. 


\begin{tcolorbox}[colback=myCol,breakable]
Recommendations:
\begin{itemizePacked}
\item Develop reliable indicators for assessing the compliance of combustion models with underlying CFD-solution.
\item Consider criteria that indicate when models are employed outside its intended validity and violate intrinsic model assumptions.
\item Develop hierarchical modeling strategies to meet user-specific requirements on accuracy and model fidelity in predicting specific response data of interest.
\item Formulate closure models that employ consistent principles for the representation of turbulence/chemistry interaction, turbulent scalar fluxes, and turbulent stresses.
\item Ensure and demonstrate that a model extension is a proper superset of the underlying parent model.
\item Develop subgrid-closure models for turbulent transport, turbulence/chemistry coupling, and turbulence/radiation interaction and other unclosed terms that are based on fundamental compressible reacting flow-theory instead of relying on non-reacting and isothermal flow models.
\end{itemizePacked}
\end{tcolorbox}
\section{\label{SEC_NUM_DISCR}What is the impact of the numerical discretization on the combustion simulation?}
In simulations of turbulent reacting flows, it is difficult to separate numerical errors from modeling errors~\cite{POPE_NJP2004}. This is attributed to the overlap of mesh resolution and physical scales of the flow as well as the non-linear interaction of physical models and filters with the numerical discretization scheme~\cite{Klein2008,KAUL_RAMAN_BALARAC_PITSCH_POF2009}. Commonly employed in simulations of turbulent reacting flows are low-order schemes that utilize 
finite-difference (FD) or finite-volume (FV) discretizations. However, by representing complex geometries that require unstructured and skewed meshes, these schemes can introduce appreciable amounts of numerical dissipation and dispersion errors, which can deteriorate the representation of the flame structure, turbulent stresses, transport, and turbulence/chemistry coupling. Mitigating the role of discretization errors is therefore critical to ensure the general behavior of a combustion model that is based on physical principles. Explicitly filtered LES-methods enable the formal separation of numerical discretization and modeling errors. Success of these methods has been demonstrated in \emph{a priori} and \emph{a posteriori} studies~\cite{RADHAKRISHNAN_BELLAN_JFM2012,COCKS20153394,GALLAGHER_SANKARAN_AIAAJ2019}. More recently, coupling effects between numerical discretization and filtering on numerical errors were investigated~\cite{EDOH_GALLAGHER_CTR2019}. Different spatial discretization schemes were examined in the context of explicitly filtered LES, showing that increasing the filter-to-grid ratio reduces the impact of numerical errors on the simulation results. 

A common technique for mitigating discretization errors is mesh adaptation. Perhaps the simplest approach is to employ local and static mesh stretching. In this method, the mesh is locally stretched or contracted to either reduce or refine the resolution in certain regions of interest. Mesh stretching can be accomplished through directional refinement for structured discretizations, and the order of accuracy can be preserved through mapping onto a regular reference grid. Unstructured solution methods are not constrained by directional stretching, and provide greater flexibility in locally adapting to complex flow features and complex flow geometries. 

Other approaches for mesh adaptation are nested grids or dynamically adaptive mesh refinement (AMR). In nested-grid techniques, which has been employed in weather forecasting~\cite{GIORGI_MEARNS_JGRA1999},  one or multiple finer meshes are statically embedded in a background mesh. To increase the physical fidelity, different submodels and additional physical representations can be utilized in the nested region. This approach enables communication of coarse-scale simulations with more refined mesoscale formulations. Coupling between the outer and nested domains is implemented either through one-way coupling, in which only the outer domain is providing boundary conditions to the nested region, or two-way coupling, in which both domains are tightly coupled through the exchange of boundary conditions. With relevance to embedding, it is important to recognize that the accuracy and performance of a nested-grid method is inherently dependent on constraints and matching conditions at the static boundaries. Another issue is to enforce conservation properties for mass, momentum, species, and energy across domain interfaces. Although the lack of conservation can affect the simulation accuracy, these methods are commonly employed for short-term predictions that do not rely on exact conservation properties. Nested-grid methods with two-way coupling can be considered as a special case of static non-uniform meshing. In contrast to nested-grid techniques with heterogeneous model description, AMR is based on a dynamic mesh adaptation in which different patch- or block-based refinement regions are represented using the same discretization and physical submodels~\cite{BERGER_COLELLA_JCP1989,BELL_BERGER_SALTZMAN_WECLOME_SIAMJ1994,AMR_BOOK2005}. For steady-state flows, AMR reduces to the nested grid approach. AMR has been employed for simulating shock waves, flow-discontinuities, and large-scale flames~\cite{GAO_GROTH_JCP2010,HILL_PULLIN_JCP2004,BELL_ETAL_PNAS2005,AMROC,AMREX,LAPOINTE_ETAL_FSJ2020}. 

Apart from mesh-refinement, the adaptation of the solution-representation offers another approach to resolve flow-field features of importance. High-order methods, such as the class of spectral and discontinuous Galerkin (DG) schemes, are techniques that -- in addition to mesh-adaptation ($h$-adaptation) -- enable refinement in polynomial order ($p$-refinement). These variational discretization techniques provide opportunities for mitigating numerical approximation errors by enabling the separation between numerical discretization and physical modeling. Specifically, compared to conventional FV/FD-schemes, advantages of DG-methods are that they (i) provide high-order accuracy on unstructured grids and complex geometries, (ii) are well suited for advanced refinement strategies, using local mesh adaptation and refinement in polynomial order, (iii) enable a compact discretization with subcell resolution, (iv) have low numerical dissipation and dispersion, and (v) the large degree of structured computations and data locality introduce a high level of parallelism, making these methods particularly suitable for high-performance computing. Since variational methods employ an element-local discretization, they are suitable for unstructured meshes, and can therefore be combined with local mesh refinement. Rigorous convergence proofs for DG-methods have been established, showing optimal convergence $\epsilon \sim h^{p+1}$ for general meshes, and the solution accuracy does not deteriorate with element type. 
\nomenclature{$\epsilon$}{error}
\nomenclature{$h$}{mesh resolution}
\nomenclature{$p$}{polynomial order}

DG-methods support local and flexible $hp$-strategies, which offers greater flexibility compared to conventional $h$-adaptation via nesting, AMR, or grid stretching alone. Theory and recent computational results have shown that for turbulent flows at low and moderate Mach numbers, an increase in polynomial order ($p$-refinement) is more effective in capturing turbulence characteristics, while for highly compressible turbulence regimes, $h$-refinement becomes beneficial for representing shows and flow-discontinuities. Significant progress has been made on extending DG-methods for reacting turbulent flows~\cite{LV_IHME_JCP2014,LV_IHME_PCI35,LV_IHME_AMS2017,WU_MA_LV_IHME_AIAA2017}. However, in order to fully utilize the potential of these high-order variational methods for  simulating turbulent flows~\cite{WANG_ETAL_IJNMF2013,GASSNER_BECK_TCFD2013,DEWIART_HILLEWAERT_ETAL_IJNMF2015,LV_MA_IHME_CF2018}, open research issues remain regarding the formulation of stabilization techniques, the development of combustion-physical models, and the construction of subgrid-closures that are consistent with the high-order discretization for large-eddy simulations.

\begin{tcolorbox}[colback=myCol,breakable]
Recommendations:
\begin{itemizePacked}
\item Develop methods to characterize and separate contributions of numerical discretization errors, physical models, and filtering operators.
\item Develop novel analytical methods for examining the impact of numerical discretization for non-linear PDEs.
\item Establish robust measures to quantify mesh dependence on numerical simulations.
\item Explore and quantify the merit of high-order discretization techniques for application to turbulent reacting flows.
\end{itemizePacked}
\end{tcolorbox}
\section{\label{SEC_MODEL_ACCURACY}How to assess the simulation accuracy?}
Directly connected to the performance evaluation of a combustion model is the quantification of the agreement with measurements and between models. So far, such comparisons have been largely performed by considering individual scalar quantities. These evaluations follow conventional statistical analysis in which moments (typically mean and root-mean-square) and conditional data are compared along axial and radial locations in the flame. Qualitative comparisons of scatter data are commonly performed to examine as to whether a particular combustion model is able to capture certain combustion-physical events such as reignition, extinction, or the departure from thermochemical equilibrium.

Another issue towards the quantitative comparison of simulation results with measurements is the multitude of experimental techniques for data acquisition, which include single-point data, line measurements, line-of-sight absorption, or multidimensional imaging at acquisition rates ranging from single-shot to high-repetition rate measurements to resolve turbulent dynamics~\cite{ALDEN_BOOD_LO_RICHTER_PCI2011}. This data is then processed in the form of statistical results from Favre and Reynolds averaging, conditional data, probability density functions, and scatter data.

Different methods have been proposed to measure model errors, and most common is the linear scalarization and comparison of single-point statistics. Since thermochemical species in chemically reacting flows are strongly coupled, the comparison of individual quantities neglects these interdependencies, often showing that major species and temperature exhibit good qualitative agreement while sensitive quantities and minor species exhibit significant deficiencies. As such, the comparison of individual scalar quantities limits a holistic examination of interscalar dependencies and the identification of correlations between reacting and hydrodynamic flow-field quantities. This has the consequence that these comparisons often only provide an inconclusive assessment of the model performance, and limit a quantitative comparison among different modeling approaches. 

Recently, the Wasserstein metric was introduced as a generalized measure for the quantitative evaluation of combustion models~\cite{JOHNSON_WU_IHME_CF2017}. Compared to commonly employed techniques that consider low-order statistical moments, this probability metric is formulated in distribution space, thereby enabling the direct consideration of instantaneous data that are obtained from transient simulations and high-speed measurements without the need for data reduction to low-order statistical moments. The Wasserstein metric and related probabilistic measures~\cite{GIBBS_SU_ISR2002}, such as the earth mover's distance, Kullback-Leibler divergence, or the Kolmogorov metric, represent frameworks for combining different data that originate  from scatter plots, instantaneous simulation results, or the reconstruction of statistical moments in the form of empirical distributions. Compared to other metrics, an attractive feature of the Wasserstein metric is that it is equipped with essential properties of metric spaces. 
\nomenclature{$W_2$}{Wasserstein metric}

\begin{figure}[!htb!]
\centering
\includegraphics[width=0.55\textwidth, clip=, keepaspectratio]{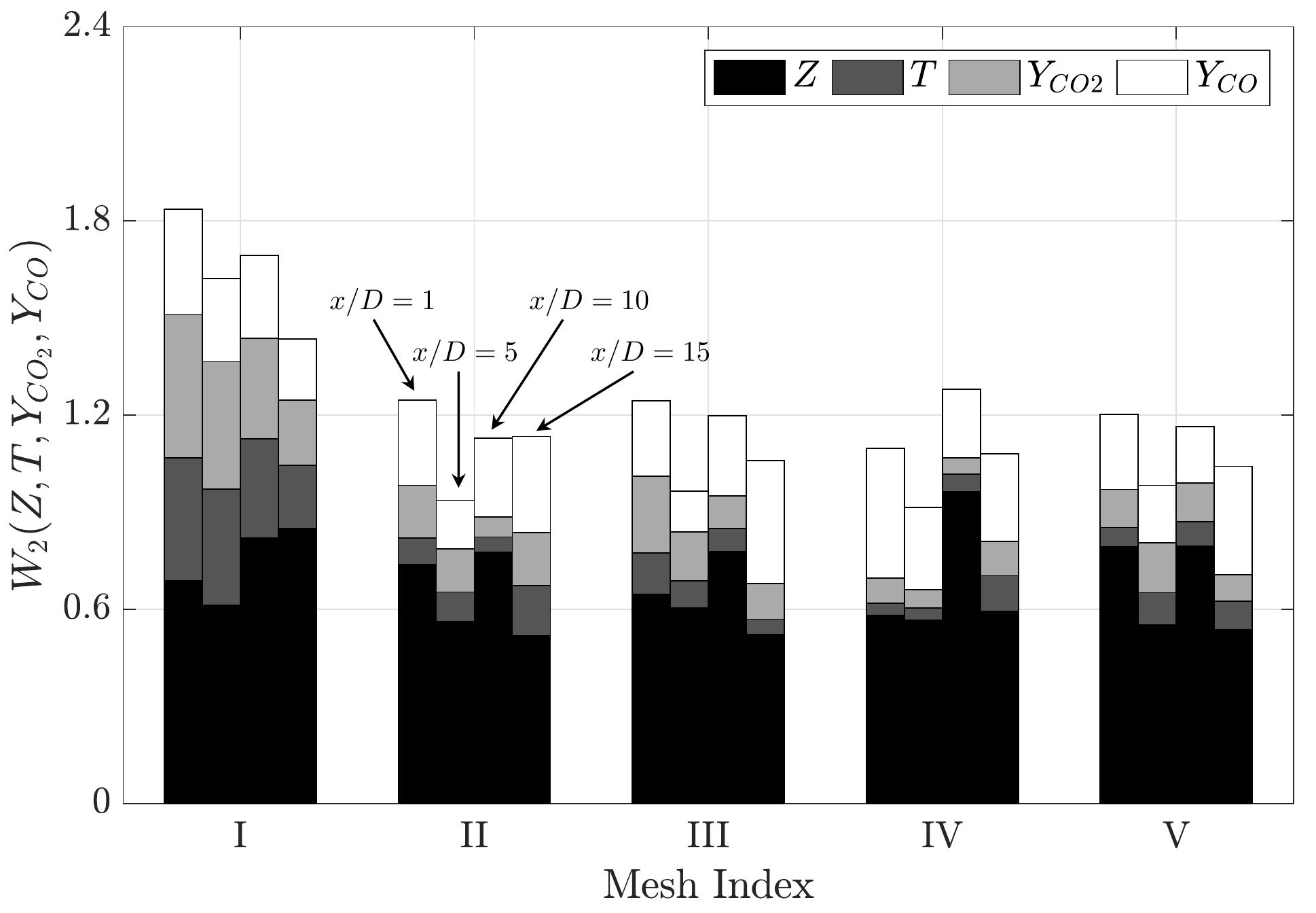}
\caption{\label{FIG_WSM}Quantitative evaluation of multiscalar Wasserstein metric~\cite{JOHNSON_WU_IHME_CF2017}, $W_2(Z,T,Y_{\ce{CO2}}, Y_{\ce{CO}})$ LES-calculations of the piloted turbulent \ce{CH4}/air jet flame with inhomogeneous inlets~\cite{BARLOW_MEARES_MAGNOTTI_CUTCHER_MASRI_CF2015} as a function of successive mesh refinement~\cite{WANG_IHME_CF2019} (Mesh I: $26\times10^5$, Mesh II: $207\times10^5$, Mesh III: $1.66\times10^6$, Mesh IV: $13.2\times10^6$, and Mesh V: $53\times10^6$ control volumes). The four bar-graphs from each contribution correspond to axial locations of $x/D=\{1,5,10,15\}$. This quantitative validation analysis enables models to be compared objectively.}
\end{figure}

The Wasserstein metric has been employed to assess different modeling approaches for simulating turbulent flames~\cite{JOHNSON_WU_IHME_CF2017,WU_MA_LV_IHME_AIAA2017,BARLOWTNF2018}. Representative results are illustrated in~\cref{FIG_WSM}, showing comparisons of the multiscalar Wasserstein metric for QoIs of mixture fraction, temperature, and species mass fractions of \ce{CO2} and \ce{CO} in simulations of a turbulent \ce{CH4}/air jet flame with inhomogeneous inlets. In this LES study, a flamelet model in conjunction with a regularized deconvolution model was employed to examine the impact of successive mesh-refinement on convergence~\cite{WANG_IHME_CF2019}. Convergence analysis of the Wasserstein metric with respect to mesh refinement provides a quantitative evaluation of the performance of a particular combustion  model and the impact of subgrid closures. Different granularities of this metric in the form of planar data, axial comparisons, and radial profiles allow for systematically assessing simulation results in predicting user-specific QoIs. As such, the Wasserstein metric allows for the quantitative evaluation of the influence of boundary conditions and mesh resolution, for isolating regions of physical complexity that demand further experimental probing, for identifying model deficiencies, and for guiding the selection of modeling strategies to meet user-specific requirements on simulation accuracy and computational cost.

While the Wasserstein metric targets assessing the accuracy in predicting QoIs, metrics for evaluating the accuracy in predicting transient combustion processes requires the consideration of the dynamics of a simulation. To this end, the Lyapunov exponent has been suggested as a possible metric for assessing the chaotic dynamics of LES calculations~\cite{NASTAC_LABAHN_MAGRI_IHME_PRF2017}. The Lyapunov exponent, $\lambda$, is amenable to a simple physical interpretation: If a system is chaotic, given an infinitesimal initial perturbation to the solution, two trajectories of the system separate in time exponentially until nonlinear saturation. The average exponential separation is the Lyapunov exponent. As such, the Lyapunov exponent provides a convenient measure of the dynamic nature in turbulent flows, and its reciprocal is related to the predictability horizon of a chaotic solution.  The Lyapunov exponent is  (i) a robust indicator of chaos, (ii)  a global quantity describing the strange attractors, and (iii) relatively simple to calculate~\cite{NoninfinitesimalPerturbations,Boffetta2002}.  The Lyapunov exponent can be employed in transient simulations where a statistically stationary flow is not present. The calculation of the Lyapunov exponent can be performed on arbitrary meshes or geometries and incorporates information about the numerical discretization, combustion models, and contains non-local information about the turbulence and boundary conditions. Unlike the Wasserstein metric, a limitation of the Lyapunov exponent is that it does not exhibit metric properties, so that theoretical information about asymptotic limits and convergence rates is not available and specific to particular applications of interest.
\nomenclature{$\lambda$}{Lyapunov exponent}

\begin{tcolorbox}[colback=myCol,breakable]
Recommendations:
\begin{itemizePacked}
\item Quantitative methods for assessing the accuracy of simulation results are needed. The  Wasserstein metric and related probabilistic measures offer compact representations of model performance at different levels of granularity for objectively analysis of simulation results. Other methods that improve the fidelity, robustness, and convergence are desired that broaden the usability of these analysis methods. 
\item The Lyapunov exponent provides a convenient method for evaluating the dynamic content and predictability horizon of a simulation; further developments on quantifying the accuracy in assessing the prediction of dynamical processes are needed that overcome deficiencies of global exponents.
\item The development of methods are necessary that address the lack of measures for the quantitative assessment of time-dependent processes; methods such as proper orthogonal decomposition, dynamic mode decomposition, or Koopman operator provide viable directions for combustion applications~\cite{HOLMES_LUMLEY_BERKOOZ_ROWLEY_BOOK2012,SCHMID_JFM2010,TAIRA_ETAL_AIAAJ2017,ROWLEY_DAWSON_ARFM2017}.
\end{itemizePacked}
\end{tcolorbox}
\section{\label{SEC_AUG_SIM}How can simulations be augmented with experimental data?}
The quantitative prediction of user-specific quantities and processes of interest within prescribed accuracy requirements introduces significant challenges. These challenges arise from the chaotic nature of turbulent reacting flow systems, uncertainties in thermochemical properties, reaction rates, constitutive relations, incomplete specifications of boundary and initial conditions, numerical errors that are introduced through the discretization, and closure models to represent the turbulence/chemistry interaction, turbulent scalar fluxes, and subgrid contributions  that evolve at the unresolved scales. While progress in any of these aspects will have a direct impact on our ability to predict turbulent flows, enormous advances and innovative approaches are required to resolve all of these aspects. 

The enrichment of simulations with data either from experiments or detailed simulations can hereby help to constrain deficiencies of current simulation approaches. In particular, remarkable advances on the development and utilization of multi-dimensional high-speed measurement techniques has provided quantitative information about instantaneous velocity fields, temperature, and species at acquisition rates and spatial resolutions that are adequate for resolving relevant turbulent scales. Although these measurements have been used in evaluating combustion models, this has been done in the context of \emph{a posteriori} comparisons of statistical results.

Data assimilation provides interesting opportunities for integrating measurements into numerical simulations~\cite{EVENSEN_BOOK2009}, and has  been used extensively to produce initial conditions for weather prediction models and to provide representations of the spatio-temporally evolving atmospheric state~\cite{Kalnay2003}. Assimilation techniques enable the estimation of the state of a complex, dynamical system by combining incomplete and sparse experimental data with erroneous models. Data assimilation can spread information from observations in space and time to unobserved state quantities, filter the effect of random observation noise from state estimates, provide estimates of observation and model errors, and determine unknown or uncertain model parameters. 

Different assimilation techniques have been developed~\cite{ASCH_BOCQUET_NODET_BOOK2016,LABAHN_WU_HARRIS_CORITONE_FRANK_IHME_FTC2020}, they can be categorized into variational techniques such as nudging, adjoint methods, and 3D/4D-Var, and statistical techniques such as optimal interpolation and the class of Kalman-filtering methods. While variational techniques rely on optimal control theory to minimize a deterministic cost function, statistical techniques incorporate stochastic information in the form of probability distributions of model uncertainties and observation errors. In particular, the ensemble Kalman filter offers advantages for incorporating quantities not contained in the solution vector, the applicability to large-scale problems, and the robust evaluation of the error covariance.

\begin{figure}[!htb!]
\centering
\includegraphics[width=0.8\textwidth, clip=, keepaspectratio]{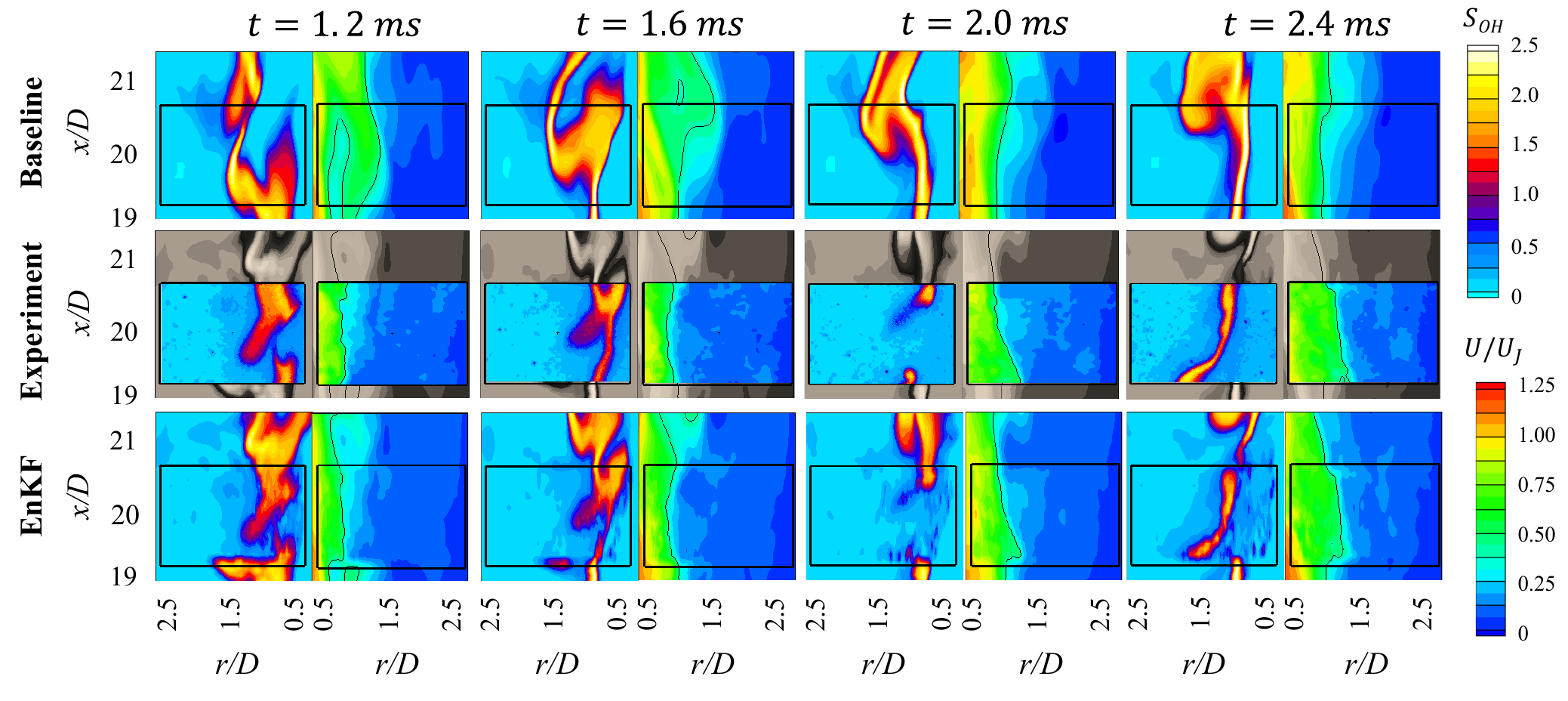}
\caption{\label{FIG_ENKF}Assimilation of simultaneous measurements into simulation of a turbulent jet flame, showing sequence of hydroxyl-LIF signal (left panel) and axial velocity (right panel) at four distinct simulation times. Top row: baseline simulation without assimilation, middle: experimental data, and bottom: results after assimilation step. Black box indicates the location of the measurement window. Figure adapted from~\cite{LABAHN_WU_CORITON_FRANK_IHME_PCI2019}.}
\end{figure}
Because of its generality, data assimilation provides interesting opportunities for turbulent-combustion applications in several ways. Specifically, assimilation can be employed for incorporating measurements into simulations under consideration of experimental uncertainties, sparsity in the data since measurements are only collected at specific locations and sample rates, and measurements are only obtained for a subset of quantities. This is illustrated in~\cref{FIG_ENKF}, where an ensemble Kalman filter (EnKF) was employed to assimilate simultaneous tomographic PIV/OH-PLIF measurements into LES with the goal of capturing local extinction and reignition events~\cite{LABAHN_WU_CORITON_FRANK_IHME_PCI2019}. While the application of data assimilation for state estimation is a common approach for atmospheric-flow predictions, forecasting has only limited value for combustion simulations. The main reasons for this are offline measurements and the current lack of real-time simulations and analysis tools. More specifically, weather-forecasting is concerned with the prediction of changes in the weather over several hours and significant efforts have been made to achieve real-time simulations. In contrast, reacting flows evolve on significantly shorter time-scales on the order of few milliseconds, and it remains illusive to capture this time horizon in real-time using existing LES-modeling capabilities. Nevertheless, state estimations can be employed for complementing experimental observations with simulations to uncover combustion-physical processes or for spinning up numerical simulations. Another attractive opportunity is to utilize data assimilation for parameter estimations and model evaluation. The merit of this concept was demonstrated in examining the performance of a flamelet-based combustion model in capturing extinction events~\cite{LABAHN_WU_CORITON_FRANK_IHME_PCI2019}, showing that the model was required to significantly reduce the mixing rate and attenuate the reactivity in order to reproduce the experimentally observed behavior of the local extinction and reignition sequence (see~\cref{FIG_ENKF}).

\begin{tcolorbox}[colback=myCol,breakable]
Recommendations:
\begin{itemizePacked}
\item Explore data assimilate techniques to integrate experimental measurements into simulations for state estimation, parameter estimations, and model evaluations.
\item Extend data assimilation methods to turbulent combustion applications by addressing aspects on scalar boundedness, ensemble density, and non-Gaussianity. 
\item Utilize experimental data and simulations for physics-informed learning, to enhance physical models, and to support model classification.
\end{itemizePacked}
\end{tcolorbox}
\section{\label{SEC_CONCL}Summary and Conclusions}
This paper examines requirements towards improving the quantitative capability of current simulation and analysis tools for predicting turbulent reacting flows. Throughout this paper, attempts are made to stipulate further development by offering recommendation for further developments. To this end,  aspects are identified that are concerned with the selection of specific simulation approaches and combustion models to achieve reliable predictions of response function of interest to the user under consideration of requirements on accuracy and computational resource allocation. Aspects on the interaction between numerical discretization and physical models are addressed; high-order methods and explicitly filtered LES strategies are discussed as promising methods for mitigating and controlling numerical discretization errors and for separating the impact of numerical discretizations from physical modeling. Related to the evaluation of numerical models, we introduce probabilistic and dynamic analysis tools in the form of the Wasserstein metric and the Lyapunov exponent, respectively, in order to support the quantitative analysis of solution accuracy and the estimation of the predictability horizon and dynamic behavior in unsteady combustion simulations; we highlight limitations of dynamic metrics in quantifying transient combustion processes. With the increasing availability of high-resolution, time-resolved, and simultaneous measurements, we examine emerging opportunities for utilizing this data to improve computational models, to obtain reliable state estimates, and to identify model deficiencies.
\section*{Acknowledgments}
Acknowledgment is made to NASA (award NX15AV04A), AFOSR (Award FA9300-19-P-1502), and the donors of The American Chemical Society Petroleum Research Fund for support of this research. Helpful discussions with Hao Wu, Peter Ma, Qing Wang, Jeff Labahn, and Alan Kerstein are gratefully acknowledged.

\end{document}